# EXPLORING EXTREME SPACE WEATHER FACTORS OF EXOPLANETARY HABITABILITY

V. S. Airapetian (GSFC/SEEC and American University, DC), V. Adibekyan (UPORTO), M. Ansdell (University of Hawaii), O. Cohen (University of Massachusetts Lowell), M. Cuntz (UFZ), W. Danchi (GSFC/SEEC), C. F. Dong (Princeton University), J. J. Drake (Harvard-Cfa), A. Fahrenbach (Caltech), K. France (CU Boulder), K. Garcia-Sage (GSFC/SEEC/CUA), A. Glocer (GSFC/SEEC), J. L. Grenfell (German Aerospace Center), G. Gronoff (NASA LaRC/SSAI), H. Hartnett (Arizona State University), W. Henning (NASA GSFC), N. R. Hinkel (Arizona State University), A. G. Jensen (UNK), M. Jin (LMSAL), P. Kalas (UC Berkeley), S. R. Kane (UC Riverside), K. Kobayashi (Yokohama National University), R. Kopparapu (GSFC/SEEC/UMD), J. Leake (GSFC), M. López-Puertas (Inst de Astrofisica de Andalucía), T. Lueftinger (University of Vienna), B. Lynch (UC Berkeley), W. Lyra (Max Planck Institute for Astronomy), A. M. Mandell (GSFC/SEEC), K. E. Mandt ( Johns Hopkins University APL), W. B. Moore (Hampton University and National Institute of Aerospace), D. Nna-Mvondo (NASA GSFC/USRA), Y. Notsu (Kyoto University), H.Maehara (Kyoto University), Y.Yamashiki (Kyoto University), K. Shibata (Kyoto University), L. D. Oman (GSFC/SEEC), R. A. Osten (STScI, JHU), A. Pavlov (GSFC/SEEC), R. M. Ramirez (ELSI), S. Rugheimer (University of St Andrews), J. E. Schlieder (GSFC/SEEC),  J. D. Schnittman (NASA GSFC), E. L. Shock (Arizona State University), C. Sousa-Silva (Massachusetts Institute of Technology), M. J. Way (NASA GISS), Y. Yang (NASA GSFC), P. A. Young (Arizona State University), G. P. Zank (University of Alabama)

*Corresponding author:
Vladimir Airapetian, NASA GSFC and American University, DC
email: vladimir.airapetian@nasa.gov

## 1. Introduction.

It is currently unknown how common life is on exoplanets, or how long planets can remain viable for life. To date, we have a superficial notion of habitability, a necessary first step, but so far lacking an understanding of the detailed interaction between stars and planets over geological timescales, dynamical evolution of planetary systems, and atmospheric evolution on planets in other systems. A planet's mass, net insolation, and atmospheric composition alone are insufficient to determine the probability that life on a planet could arise or be detected. The latter set of planetary considerations, among others, underpin the concept of the habitable zone (HZ), defined as the circumstellar region where standing bodies of liquid water could be supported on the surface of a rocky planet (e.g. [1]). However, stars within the same spectral class are *often* treated in the same way in HZ studies, without any regard for variations in activity among individual stars (e.g. [1,2]). Such formulations ignore differences in how non-thermal emission and magnetic energy of transient events in different stars affect the ability of an exoplanet to retain its atmosphere ([3-5]).

In the last few years there has been a growing appreciation that the atmospheric chemistry, and even retention of an atmosphere in many cases, depends critically on the high-energy radiation and particle environments around these stars [6-15]. Indeed, recent studies have shown stellar activity and the extreme space weather, such as that created by the frequent flares and coronal mass ejections (CMEs) from the active stars and young Sun, may have profoundly affected the chemistry and climate and thus habitability of the early Earth and terrestrial type exoplanets [3,4,11-23].

The goal of this white paper is to identify and describe promising key research goals to aid the field of the exoplanetary habitability for the next 20 years.

## 2. The key questions for exoplanet science and research.

Atmospheres of Earth-sized planets within respective HZs could be vulnerable to the high XUV fluxes from stellar flares, coronal mass ejections and dense and fast stellar winds. Atmospheres of close-in terrestrial planets around low luminosity M dwarfs could be significantly eroded within a few to hundred Myr, assuming minimal replenishment from volcanic outgassing and bombardment by comets [5,7,9,10,18]..

The study of the atmospheric evolution of Mars can provide clues to the conditions of habitability of early Mars and exoplanets. The Martian atmosphere over the last 4.5 billion years is affected by the processes of atmospheric escape, outgassing and reaction of atmospheric constituents with the surface (e.g. carbonate precipitation, [24-25]). Abundant morphological and mineralogical evidence suggests that early Mars (~ 3.8 Ga) possessed episodically - if not persistently - warmer and wetter conditions [26,27]. This most likely would have required it to have a thicker atmosphere with efficient greenhouse-induced warming (e.g., [28-31]), although other solutions have also been proposed (e.g., [32]. How Mars lost that thick atmosphere, which species were lost, and the composition and climate prior to the atmospheric loss, are all open questions that are fundamentally important for understanding the factors influencing habitability in both the evolving solar system as well as the growing population of exoplanetary worlds.

Our Exoplanetary Space Weather (ESW) community that includes Sellers Exoplanetary Environment Collaboration (SEEC) at NASA GSFC, Nexus for Exoplanet System Science (NExSS), and broader US and international members has recently developed the vision and the roadmap for this new field of astrobiology as the product of the "ESW, Climate and Habitability" workshop [33]. We concluded that the evolving star-exoplanetary system is critically important for developing conditions supporting life, and thus should be considered as an important aspect of strategies to search for and detect life in the Universe..

The proposed key questions to address over the next 20 years relate to the origin of ESW, its impact on atmospheric erosion and chemistry of terrestrial planets, the conditions initiating prebiotic chemistry, developing biogenic zones and atmospheric biosignatures that can be addressed using a broad range of multi-dimensional multi-fluid physico-chemical atmospheric models. These models should be validated for the extreme conditions in our solar system along with computational capabilities, laboratory experiments, and development of mission concepts.

Star-planet interactions should be modeled in a global planetary system environment with a systematic, integrated approach using theoretical and observational methods combining tools and methodologies of the four NASA science disciplines: astrophysics, heliophysics and planetary and Earth science.

### 2.1. Exploring Evolution of Exoplanetary Atmospheres

The long-term evolution of the exoplanetary atmospheres is a result of the interplay between solar radiation, particle, exogenic and endogenic forces that are involved in modifying the planet's environment on a scale of billions of years. Research in the upcoming decade must address this major question as it affects the physico-chemical atmospheric changes on rocky planets and on its effects on climate.

One crucial question for exoplanetary habitability is: *How is ESW important in specifying the rate of atmospheric escape of neutral and ion species and its effects on habitability of exoplanets with different surface gravities?* In order to comprehensively address this question, the community needs to consider the entire space weather environment - including the planetary thermosphere (about 100 km) through the magnetosphere. We recommend modifications of existing space weather models for exoplanetary environments to calculate the magnetospheric response and associated atmospheric heating and escape rates. Global magnetospheric models coupled with a kinetic model of the ring current and inner plasma sheet and their coupling to the ionosphere, provide an appropriate magnetospheric response that can then be fed into a global ionosphere-thermosphere-mesosphere (ITM) model to provide the temperature and pressure, the resultant atmospheric chemistry and atmospheric escape rates. Such coupled code suites have already been developed as a part of Community Coordinated Modeling Center at NASA GSFC [34].

Providing appropriate stellar boundary conditions is an integral part of the recommended approach. Significant observational data for K-M dwarfs is available (e.g., [35-38]), but is incomplete in age, spectral type, and X-ray and UV (XUV) emission. Our recommended program should combine advanced numerical and observational approaches. XUV spectroscopic data and direct measurements of stellar surface magnetic fields should be used as direct inputs for heliophysics-based multidimensional models of stellar coronae, winds and CMEs [see for example, [39]). These results will provide high fidelity inputs to numerical models of the thermodynamic and chemical evolution of exoplanet atmospheres, leading to knowledge of the habitable conditions of an exoplanet over time. These models will specify the initial and boundary conditions for the coupled ionosphere-thermosphere models that describe the ion and neutral temperature and density of the coupled ionosphere-thermosphere systems [18,22]. These upper atmosphere models will provide the realistic assessment of hydrodynamic, ion and neutral escape from atmospheres of exoplanets around K-M dwarfs.

An intriguing aspect of the interaction between ESW and atmospheres relates to the creation and evolution of large abiotically-generated oxygen atmospheres [10; Harman et al. 2015). These atmospheres can be produced during the extended pre-main sequence phase predicted for the lowest mass stars (e.g. Baraffe et al.). During this phase, and for the first few Gyr after the star begins fusing hydrogen, the star is very active and flares could blow off the atmosphere. Alternatively, ionizing radiation (i.e., UV and XUV) could drive vigorous photochemistry and detectable auroral features (e.g.,[40,41]). These atmospheres are not present in our Solar System, but they may be common in HZs around K and M dwarfs, i.e. precisely the planets that will be discovered by TESS and remotely probed by JWST transit transmission and eclipse spectroscopy. A clear understanding of the evolution of these atmospheres in the context of ESW will be vital for the successful interpretation of JWST spectra [42].

**2.2. Exploring Exoplanetary Biogenic Zones**

The *Kepler* Space Telescope, in its mission to discover exoplanets, has characterized frequent and energetic flares (or superflares) on solar-type stars, providing a mechanism by which host stars may have profound effects on the physical and chemical evolution of exoplanetary atmospheres [43]. Kepler survey provided clear correlation between maximum flare energy and stellar activities represented by starspot area that can be extended to predict possible maximum impacts of stellar energetic particle events (SEPs) for individual exoplanets from their host stars. These effects may represent a vastly underestimated factor regarding the chemical impacts of energetic particles including SEPs and the relative contribution from Galactic Cosmic Rays. The Earth's middle atmospheric chemistry during large SEPs is represented in production of $NO_x$, $HO_x$ constituents and destruction of ozone [19,44,45]. These studies aim to provide answers to the following three crucial questions: (1) What are the strongest biosignatures of an Earth-like biosphere which experiences XUV/SEP fluxes? (2) How significant is ozone destruction in oxygenated exoplanetary atmospheres of Earth-like planets around G-M dwarfs? (3) What are the effects of superflare-induced SEPs on the prebiotic chemistry of rocky exoplanets around G-M dwarfs - and at what level can the particle radiation be lethal for surface life forms?

Life requires the formation of biologically important molecules that can create chemical disequilibrium by forming complex molecules from simpler species that ultimately produced life forms - the process known as abiogenesis. This rise of complexity could be driven as a by-product of persistent energy sources including UV radiation, shock heating from impacts, electric discharges, radioactivity and solar and cosmic ray particles [17,46-49]. The crucial raw atmospheric molecular ingredients of life as we know it include $N_2$, $CO_2$, $CH_4$ and $H_2O$. The definition of habitable zone has recently been enhanced to a biogenic zone (BZ), within which the stellar ionizing energy fluxes are efficient enough to ignite reactive chemistry through ionization and dissociation of these molecular ingredients [17]. Hydrogen cyanide, the major product, mixes with water vapor and forms formamide that may serve as a prerequisite for the complex molecules crucial for life [50,51]. This chemistry also forms $N_2O$, a powerful greenhouse gas and a possible biosignature, that may keep the atmosphere warm for liquid surface water to exist. Recent studies of chemical impacts of exoplanetary atmospheres around active stars show the efficient production of $NO_x$ and $OH_x$ which scale linearly with the particle fluxes [19,52,53]. These studies are providing necessary context for the future detection of Earth-like biospheres' signatures around active stars.

3. **Recommendations:**

To make significant progress in understanding the atmospheric factors and their effects on (exo)planetary surface habitability, priority should be given to the following:
1. Observational studies of the variability of activity on K-M dwarfs and predictive models of superflares and CME initiation, including the formation of energetic particle and high energy photon fluxes (especially XUV) on well resolved evolutionary timescales and comprehension of average and maximum flare energy of those dwarfs;
2. Enhanced system modeling of the links between stellar evolution, stellar activity (including the production of ionizing radiation and energetic particles;
3. Multi-dimensional modeling efforts of magnetosphere-ionosphere-thermosphere-mesosphere system and atmospheric erosion resulting from various modes of star-planet interactions;
4. Multi-dimensional climate models of exoplanetary atmospheres modified by ESW.
5. Theoretical and laboratory studies of the initiation of prebiotic chemistry and biochemical pathways to the building blocks of life;
6. Adequate (experimental or theoretical) measurements of molecular spectra for each, potentially detectable, relevant biosignature gas. This is important because the interpretation of atmospheric data requires observations to be replicated with models, which in turn require a correct and complete set of molecular spectra [54].
7. Improvement of our understanding of surface catalysis reactions involving gas-phase species;

8. Modeling and laboratory experiments of atmospheric biosignatures of life guided by prebiotic chemistry and observations of Earth's upper atmosphere in response to current space weather. This implies to re-define biosignatures in the context of exoplanetary habitability.
9. Instrument development in support of direct imaging exoplanetary missions to detect atmospheric bio-signatures including characterization of host stars;
10. Multi-mission epoch-specific space-based FUV, NUV, and X-ray observations of nearby K-M stars to permit the reconstruction of activity levels as a function of spectral type and age.
11.Magnetospheric low-frequency radio observations are crucial to provide information about planetary magnetic field, the factor of habitability [55].